# Improving Software Effort Estimation Using Neuro-Fuzzy Model with SEER-SEM

Wei Lin Du[1], Danny Ho[2], Luiz Fernando Capretz [3]

*Abstract* - Accurate software development effort estimation is a critical part of software projects. Effective development of software is based on accurate effort estimation. Although many techniques and algorithmic models have been developed and implemented by practitioners, accurate software development effort prediction is still a challenging endeavor in the field of software engineering, especially in handling uncertain and imprecise inputs and collinear characteristics. In order to address these issues, previous researchers developed and evaluated a novel soft computing framework. The aims of our research are to evaluate the prediction performance of the proposed neuro-fuzzy model with System Evaluation and Estimation of Resource Software Estimation Model (SEER-SEM) in software estimation practices and to apply the proposed architecture that combines the neuro-fuzzy technique with different algorithmic models. In this paper, an approach combining the neuro-fuzzy technique and the SEER-SEM effort estimation algorithm is described. This proposed model possesses positive characteristics such as learning ability, decreased sensitivity, effective generalization, and knowledge integration for introducing the neuro-fuzzy technique. Moreover, continuous rating values and linguistic values can be inputs of the proposed model for avoiding the large estimation deviation among similar projects. The performance of the proposed model is accessed by designing and conducting evaluation with published projects and industrial data. The evaluation results indicate that estimation with our proposed neuro-fuzzy model containing SEER-SEM is improved in comparison with the estimation results that only use SEER-SEM algorithm. At the same time, the results of this research also demonstrate that the general neuro-fuzzy framework can function with various algorithmic models for improving the performance of software effort estimation.

## I. INTRODUCTION

The cost and delivery of software projects and the quality of products are affected by the accuracy of software effort estimation. In general, software effort estimation techniques can be subdivided into experience-based, parametric model-based, learning-oriented, dynamics-based, regression-based, and composite techniques (Boehm, Abts, and Chulani 2000). Amongst these methods, model-based estimation techniques involve the use of mathematical equations to perform software estimation. The estimation effort is a function of the number of variables, which are factors impacting software cost (Boehm 1981). These model-based estimation techniques comprise the general form: $E = a \times Size^b$, where E is the effort, size is the product size, a is the productivity parameters or factors, and b is the parameters for economies or diseconomies (Fischman, McRitchie, and Galorath 2005; Jensen, Putnam, and Roetzheim 2006). In the past decades, some important software estimation algorithmic models have been published by researchers, for instance Constructive Cost Model (COCOMO) (Boehm et al. 2000), Software Life-cycle Management (SLIM) (Putnam and Myers 1992), SEER-SEM (Galorath and Evans 2006), and Function Points (Albrecht 1979; Jones 1998). Model-based techniques have several strengths, the most prominent of which are objectivity, repeatability, the presence of supporting sensitivity analysis, and the ability to calibrate to previous experience (Boehm 1981). On the other hand, these models also have some disadvantages. One of the disadvantages of algorithmic models is their lack of flexibility in adapting to new circumstances. The new development environment usually entails a unique situation, resulting in imprecise inputs for estimation by an algorithmic model. As a rapidly changing business, the software industry often faces the issue of instability and hence algorithmic models can be quickly outdated. The outputs of algorithmic models are based on the inputs of size and the ratings of factors or variables (Boehm 1981). Hence, incorrect inputs to such models, resulting from outdated information, cause the estimation to be inaccurate. Another drawback of algorithmic models is the strong collinearity among parameters and the complex non-linear relationships between the outputs and the contributing factors.

SEER-SEM appeals to software practitioners because of its powerful estimation features. It has been developed with a combination of estimation functions for performing various estimations. Created specifically for software effort estimation, the SEER-SEM model was influenced by the frameworks of Putnam (Putnam and Myers 1992) and Doty Associates (Jensen, Putnam, and Roetzheim 2006). As one of the algorithmic estimation models, SEER-SEM has two main limitations on effort estimation. First, there are over 50 input parameters related to the various factors of a project, which increases the complexity of SEER-SEM, especially for managing the uncertainty from these outputs. Second, the specific details of SEER-SEM increase the difficulty of discovering the nonlinear relationship between the parameter inputs and the corresponding outputs. Overall, these two major limitations can lead to a lower accuracy in effort estimation by SEER-SEM.

*About[1] Wei Lin Du, the Department of Electrical and Computer Engineering, the University of Western Ontario, London, Ontario, Canada N6A 5B9*
*(email: wdu6@uwo.ca)*
*About[2] Danny Ho, NFA Estimation Inc., Richmond Hill, Ontario Canada L4C 0A2*
*(email: danny@nfa-estimation.com)*
*About[3] Dr. Luiz Fernando Capretz, the Department of Electrical and Computer Engineering, the University of Western Ontario, London, Ontario, Canada N6A 5B9*
*(telephone: 1-519-661-2111 ext. 85482 email: lcapretz@eng.uwo.ca)*

The estimation effort is a function of the number of variables, which are factors impacting software cost (Boehm 1981). These model-based estimation techniques comprise the general form: $E = a \times Size^b$, where E is the effort, size is the product size, a is the productivity parameters or factors, and b is the parameters for economies or diseconomies (Fischman, McRitchie, and Galorath 2005; Jensen, Putnam, and Roetzheim 2006). In the past decades, some important software estimation algorithmic models have been published by researchers, for instance Constructive Cost Model (COCOMO) (Boehm et al. 2000), Software Life-cycle Model (SLIM) (Putnam and Myers 1992), SEER-SEM (Galorath and Evans 2006), and Function Points (Albrecht 1979; Jones 1998). Model-based techniques have several strengths, the most prominent of which are objectivity, repeatability, the presence of supporting sensitivity analysis, and the ability to calibrate to previous experience (Boehm 1981). On the other hand, these models also have some disadvantages. One of the disadvantages of algorithmic models is their lack of flexibility in adapting to new circumstances. The new development environment usually entails a unique situation, resulting in imprecise inputs for estimation by an algorithmic model. As a rapidly changing business, the software industry often faces the issue of instability and hence algorithmic models can be quickly outdated. The outputs of algorithmic models are based on the inputs of size and the ratings of factors or variables (Boehm 1981). Hence, incorrect inputs to such models, resulting from outdated information, cause the estimation to be inaccurate. Another drawback of algorithmic models is the strong collinearity among parameters and the complex non-linear relationships between the outputs and the contributing factors.

SEER-SEM appeals to software practitioners because of its powerful estimation features. It has been developed with a combination of estimation functions for performing various estimations. Created specifically for software effort estimation, the SEER-SEM model was influenced by the frameworks of Putnam (Putnam and Myers 1992) and Doty Associates (Jensen, Putnam, and Roetzheim 2006). As one of the algorithmic estimation models, SEER-SEM has two main limitations on effort estimation. First, there are over 50 input parameters related to the various factors of a project, which increases the complexity of SEER-SEM, especially for managing the uncertainty from these outputs. Second, the specific details of SEER-SEM increase the difficulty of discovering the nonlinear relationship between the parameter inputs and the corresponding outputs. Our study attempts to reduce the negative impacts of the above major limitations of the SEER-SEM effort estimation model on prediction accuracy and make contributions towards resolving the problems caused by the disadvantages of algorithmic models. First, for accurately estimating software effort the neural network and fuzzy logic approaches are adopted to create a neuro-fuzzy model, which is subsequently combined with SEER-SEM. The Adaptive Neuro-Fuzzy Inference System (ANFIS) is used as the architecture of each neuro-fuzzy sub-model. Second, this research is another evaluation for effectiveness of the general model of neuro-fuzzy with algorithmic model proposed by the previous studies. Third, the published data and industrial project data are used to evaluate the proposed neuro-fuzzy model with SEER-SEM. Although the data was collected specifically for COCOMO 81 and COCOMO 87, they are transferred from COCOMOs to COCOMO II and then to the SEER-SEM parameter inputs, utilizing the guidelines from the University of Southern California (USC) (Madachy, Boehm, and Wu 2006; USC Center for Software Engineering 2006). After the transfer of this data, the estimation performance is verified to ensure its feasibility.

## II. Background

Soft computing, which is motivated by the characteristics of human reasoning, has been widely known and utilized since the 1960s. The overall objective from this field is to achieve the tolerance of incompleteness and to make decisions under imprecision, uncertainty, and fuzziness (Nauck, Klawonn, and Kruse 1997; Nguyen, Prasad, Walker, and Walker 2003). Because of capabilities, soft computing has been adopted by many fields, including engineering, manufacturing, science, medicine, and business. The two most prominent techniques of soft computing are neural networks and fuzzy systems. The most attractive advantage of neural networks is the ability to learn from previous examples, but it is difficult to prove that neural networks are working as expected. Neural networks are like "black boxes" to the extent that the method for obtaining the outputs is not revealed to the users (Chulani 1999; Jang, Sun, and Mizutani 1997). The obvious advantages of fuzzy logic are easy to define and understand an intuitive model by using linguistic mappings and handle imprecise information (Gray and MacDonell 1997; Jang, Sun, and Mizutani 1997). On the other hand, the drawback of this technique is that it is not easy to guarantee that a fuzzy system with a substantial number of complex rules will have a proper degree of meaningfulness (Gray and MacDonell 1997). In addition, the structure of fuzzy if-then rules lacks the adaptability to handle external changes (Jang, Sun, and Mizutani 1997). Although neural networks and fuzzy logic have obvious strengths as independent systems, their disadvantages have prompted researchers to develop a hybrid neuro-fuzzy system that minimizes these limitations. Specifically, a neuro-fuzzy system is a fuzzy system that is trained by a learning algorithm derived from the neural network theory (Nauck, Klawonn, and Kruse 1997). Jang's (Jang, Sun, and Mizutani 1997; Nauck, Klawonn, and Kruse 1997) ANFIS is one type of hybrid neuro-fuzzy system, which is composed of a five-layer feed-forward network architecture.

Soft computing is especially important in software cost estimation, particularly when dealing with uncertainty and with complex relationships between inputs and outputs. In the 1990's a soft computing technique was introduced to build software estimation models and improve prediction performance (Damiani, Jain, and Madravio 2004). As a technique containing the advantages of the neural networks and fuzzy logic, the neuro-fuzzy model was adopted for software estimation. Researchers developed some models with the neuro-fuzzy technique and demonstrated their ability to improve prediction accuracy. Hodgkinson and Garratt (*Hodgkinson and Garratt 1999)* introduced the neuro-fuzzy model for cost estimation as one of the important methodologies for developing non-algorithmic models. Their model did not use any of the existing prediction models, as the inputs are size and duration, and the output is the estimated project effort. The clear relationship between Function Points Analysis (FPA)'s primary component and effort was demonstrates by Abran and Robillard's study (Abran and Robillard 1996). Huang *et al.* (Huang, Ho, Ren, and Capretz 2005 and 2006) proposed a software effort estimation model that combines a neuro-fuzzy framework with COCOMO II. The parameter values of COCOMO II were calibrated by the neuro-fuzzy technique in order to improve its prediction accuracy. This study demonstrated that the neuro-fuzzy technique was capable of integrating numerical data and expert knowledge. And the performance of PRED(20%) and PRED(30%) were improved by more than 15% and 11% in comparison with that of COCOMO 81. Xia *et al.* (Xia, Capretz, Ho, and Ahmed 2008) developed a Function Point (FP) calibration model with the neuro-fuzzy technique, which is known as the Neuro-Fuzzy Function Point (NFFP) model. The objectives of this model are to improve the FP complexity weight systems by fuzzy logic, to calibrate the weight values of the unadjusted FP through the neural network, and to produce a calibrated FP count for more accurate measurements. Overall, the evaluation results demonstrated that the average improvement for software effort estimation accuracy is 22%. Wong *et al.* (Wong, Ho, and Capretz 2008) introduced a combination of neural networks and fuzzy logic to improve the accuracy of backfiring size estimates. In this case, the neuro-fuzzy approach was used to calibrate the conversion ratios with the objective of reducing the margin of error. The study compared the calibrated prediction model against the default conversion ratios. As a result, the calibrated ratios still presented the inverse curve relationship between the programming languages level and the SLOC/FP, and the accuracy of the size estimation experienced a small degree of improvement.

### III. A Neuro-Fuzzy SEER-SEM Model

#### A. *A General Soft Computing Framework for Software Estimation*

This section describes a general soft computing framework for software estimation, which is based on the unique architecture of the neuro-fuzzy model described in the patent US-7328202-B2 (Huang, Ho, Ren, and Capretz 2008) and was built by Huang *et al.* (Huang, Ho, Ren, and Capretz 2006). The framework is composed of inputs, a neuro-fuzzy bank, corresponding values of inputs, an algorithmic model, and outputs for effort estimation, as depicted in Fig. 1.

Among the components of the proposed framework, the neuro-fuzzy bank and the algorithmic model are the major parts of the model. The inputs are rating levels, which can be continuous values or linguistic terms such as Low, Nominal, or High. $V_1, ..., V_n$ are the non-rated values of the software estimation algorithmic model. On the other hand, $AI_0, ..., AI_m$ are the corresponding adjusted quantitative parameter values of the rating inputs, which are the inputs of the software estimation algorithmic model for estimating effort as the final output.

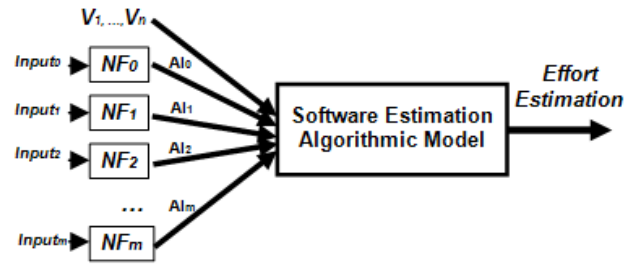

Fig.1. A General Soft Computing Framework.

This novel framework has attractive attributes, particularly the fact that it can be generalized to many different situations and can be used to create more specific models. In fact, its generalization is one of the purposes of designing this framework. Its implementation is not limited to any specific software estimation algorithmic model. The algorithmic model in the framework can be one of the current popular algorithmic models such as COCOMO, SLIM or SEER-SEM. When various algorithmic models are implemented into this framework, the inputs and the non-rating values are different.

#### B. *SEER-SEM Effort Estimation Model*

SEER-SEM stemmed from the Jensen software model in the late 1970s, where it was developed at the Hughes Aircraft Company's Space and Communications Group (Fischman, McRitchie, and Galorath 2005; Galorath and Evans 2006; Jensen, Putnam, and Roetzheim 2006). In 1988, Galorath Inc. (GAI) started developing SEER-SEM (Galorath and Evans 2006), and in 1990, GAI trademarked this model. The SEER-SEM model was motivated by Putnam's SLIM and Boehm's COCOMO (Fischman, McRitchie, and Galorath 2005; Galorath and Evans 2006; Jensen, Putnam, and Roetzheim 2006). Over the span of a decade, SEER-SEM has been developed into a powerful and sophisticated model, which contains a variety of tools for performing different estimations that are not limited to software effort. SEER-SEM includes the breakdown structures for various tasks, project life cycles, platforms, and applications. It also includes the most development languages, such as the third and fourth generation programming languages, in the estimation. Furthermore, the users can select different knowledge bases (KBs) for Platform, Application, Acquisition Method, Development Method, Development Standard, and Class based on the requirements of their projects. SEER-SEM provides the baseline settings for parameters according to the KB inputs; there are over 50 parameters that impact the estimation outputs. Among them, 34 parameters are used by SEER-SEM effort estimation model (Galorath Incorporated 2001 and 2006). Nevertheless, the SEER-SEM model contains some disadvantages. For instance, the efforts spent on pre-specification phases, such as requirements collection, are not included in the effort estimation. In SEER-SEM effort estimation, each parameter has sensitivity inputs, with the ratings ranging from Very Low (VLo-) to Extra High (EHi+). Each main rating level is divided into three sub-ratings, such as VLo-, VLo, VLo+. These ratings are translated to the corresponding quantitative value used by the effort estimation calculation. The SEER-SEM effort estimation is calculated by the following equations:

$$E = 0.393469 \times K \quad (1)$$

$$C_{tb} = 2000 \times \exp\left(\frac{-3.70945 \times \ln\left(\frac{ctbx}{4.11}\right)}{5 \times TURN}\right) \quad (2)$$

$$K = D^{0.4} \times \left(\frac{S_e}{C_{te}}\right)^{1.2}, \quad C_{te} = C_{tb}/ParmAdjustment \quad (3)$$

ctbx = ACAP × AEXPAPPL × MODP × PCAP × TOOL × TERM (4)

ParmAdjustment= LANGLEXP × TSYSTEXP × DSYSDEXP × PSYSPEXP × SIBRREUS × MULT × RDED × RLOC × DSVL × PSVL × RVOL × SPEC × TEST × QUAL × RHST(HOST) × DISP × MEMC × TIMC × RTIM × SECR × TSVL (5)

where,
$E$ is the development effort (in person years),
$K$ is the total Life-cycle effort (in person years) including development and maintenance,
$S_e$ is the Effective Size (SLOC),
$D$ is the Staffing complexity,
$C_{te}$ is the Effective technology,
$C_{tb}$ is the Basic technology.

The elements included in equations (4) and (5) are parameters or combined parameters; the formulas for calculating combined parameters are shown below:
AEXPAPPL = 0.82+(0.47*EXP(-0.95977*(AEXP/APPL))) (6)

LANGLEXP = 1+((1.11+0.085*LANG)-1)*EXP(-LEXP/(LANG/3)) (7)
TSYSTEXP = 1+(0.035+0.025*TSYS)*EXP(-3*TEXP/TSYS) (8)
DSYSDEXP = 1+(0.06+0.05*DSYS)*EXP(-3*DEXP/DSYS) (9)

PSYSPEXP
$$\begin{cases} = (0.91\wedge PSYS + 0.23*PSYS*EXP(-3*PEXP/PSYS))\wedge 0.833, \\ \quad when\ PSYS \neq 0 \\ = 1, when\ PSYS = 0 \end{cases}$$

(10)

SIBRREUS = SIBR*REUS +1

(11)

C. *A Neuro-Fuzzy Model with SEER-SEM*

a) *Overview*

This section will describe the proposed framework of the neuro-fuzzy model with SEER-SEM, based on the general structure in the section III.A, as depicted in Fig. 2. The inputs consist of two parts: non-rating inputs and the rating levels of parameters, which include 34 technology and environment parameters and 1 complexity or staffing parameter. Among the technology and environment parameters, there is one parameter (SIBR), which is not rated by the linguistic term. SIBR is decided by users, through inputting the percentage. Hence, similar to the input of size, SIBR is a non-rating value. While the other parameters are labeled as $PR_1$ to $PR_{34}$, SIBR is labeled $PR_{35}$.

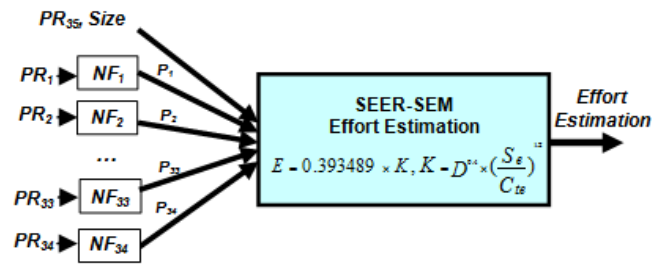

Fig.2. A Neuro-Fuzzy Model with SEER-SEM.

Each parameter PRi (i = 1, …, 34) can be a linguistic term or a continuous rating value. The linguistic inputs are from 18 rating levels (r =1, …, 18), which include Very Low– (VLo-), Very Low (VLo), Very Low+ (VLo+), Low–, Low, Low+, Nominal- (Nom-), Nominal (Nom), Nominal+ (Nom+), High – (Hi-), High (Hi), High+ (Hi+), Very High– (VHi-), Very High (VHi), Very High+ (VHi+), Extra High– (EHi-), Extra High (EHi), and Extra High+ (EHi+). In these ratings, there are 6 main levels, VLo, Low, Nom, Hi, VHi, and EHi, and each main rating level has three sub-levels: minus, plus or neutral (Galorath Incorporated 2006 be 2005). NFi (i = 1, …, 34) is a neuro-fuzzy bank, which is composed of thirty-four NFi sub-models. The rating levels of each parameter PRi (i = 1, …, 34) are the input of each NFi. Through these sub-models, the rating level of a parameter is translated into the corresponding quantitative value (Pi, i = 1, …, 34) as the inputs of the SEER-SEM effort estimation as introduced in the section III.B, from equations (1) to (11). The output of the proposed model is the software effort estimation.

    b)   *Structure of $NF_i$*

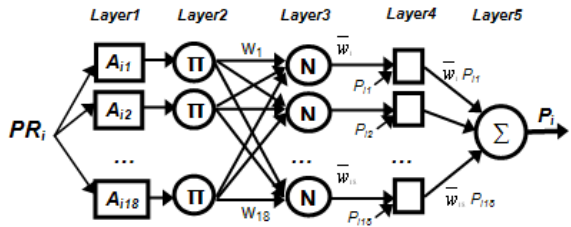

Fig.3. Structure of $NF_i$.

The neuro-fuzzy bank fulfills an important function in the proposed neuro-fuzzy model with SEER-SEM effort estimation model. $NF_i$ produces fuzzy sets and rules for training datasets. It translates the rating levels of a parameter into a quantitative value and calibrates the value by using actual project data. According to fuzzy logic techniques, linguistic terms can be presented as a fuzzy set. There are 18 rating levels for each parameter in linguistic terms, which are used to define a fuzzy set in this research. The selected membership function translates the linguistic terms in this fuzzy set to membership values. Each $NF_i$ uses the structure of the Adaptive Neuro-Fuzzy Inference System (ANFIS), which is a five-layer hybrid neuro-fuzzy system, as depicted in Fig. 3.

- **Input and Output of $NF_i$**

There is one input and one corresponding output for each NF. The input of each $NF_i$ ($PR_i$, $i = 1, \ldots, 34$) is the rating level of a parameter for SEER-SEM effort estimation model, such as Very Low (VLo) or High (Hi). On the other hand, the output is the corresponding quantitative value of this parameter ($P_i$, $i = 1, \ldots, 34$), such as 1.30.

- **Fuzzy Rule**

Based on the features of ANFIS and the structure shown in Fig. 3, this work refers to the form of the fuzzy if-then rule proposed by Takagi and Sugeno (Takagi and Sugeno 1986). The $r$th fuzzy rule of the proposed model is defined as below:

Fuzzy Rule r:  **IF** $PR_i$ is $A_{ir}$ **THEN** $P_i = P_{ir}$, $r = 1, 2, \ldots, 18$

where $A_{ir}$ is a rating level of the fuzzy set that ranges from Very Low- to Extra High+ for the $i$th parameter and is characterized by the selected membership function, and $P_{ir}$ is the corresponding quantitative value of the $r$th rating level for the $i$th parameter. Furthermore, with this fuzzy rule, the premise part is the fuzzy set and the consequent part is the non-fuzzy value. Overall, the fuzzy rules build the links between a linguistic rating level and the corresponding quantitative value of a parameter.

- **Functions of Each Layer**

*Layer 1*: In this layer, the membership function of fuzzy set A translates the input, $PR_i$, to the membership grade. The output of this layer is the membership grade of $PR_i$, which is the premise part of fuzzy rules. Also, the membership function of the nodes in this layer is utilized as the activation function; in our proposed model, all the membership functions of each node in Layer 1 are the same. In subsequent sections, the selected membership function will be discussed in detail.

$$O_r^1 = \mu_{A_{ir}}(PR_i) \quad \begin{array}{l} \text{for } i = 1, 2, \ldots, 34 \\ r = 1, 2, \ldots, 18 \end{array} \quad (12)$$

where $O_k^i$ is the membership grade of $A_{ir}$ (=VLo-, VLo, VLo+, Low-, Low, Low+, Nom-, Nom, Nom+, Hi-, Hi, Hi+, VHi-, VHi, VHi+, EHi-, EHi, or EHi+) with the input $PR_i$ or $\mu_{A_{..}}$ continuous number $x \in [0,19]$; is the membership function of $A_{ir}$.

*Layer 2*: Producing the firing strength is the primary function of this layer. The outputs of Layer 1 are the inputs of each node in this layer. In each node, Label Π multiplies all inputs to produce the outputs according to the defined fuzzy rule for this node. Consequently, the outputs of this layer are the firing strength of a rule. The premise part in the defined fuzzy rule of our proposed model is only based on one condition. Therefore, the output of this layer, the firing strength, is not changed and is thus the same as the inputs, or membership grade.

$$O_r^2 = w_r = O_r^1 = \mu_{A_{ir}}(PR_i) \quad (13)$$

*Layer 3*: The function of this layer is to normalize the firing strengths for each node. For each node, labeled "N", the ratio of the $r$th rule's firing strength to the sum of all rules' firing strengths related to PRi is calculated. The resulting outputs are known as normalized firing strengths.

$$O_r^3 = \overline{w_r} = \frac{w_r}{\sum_{r=1}^{18} w_r} \quad (14)$$

*Layer 4*: An adaptive result of $P_i$ is calculated with the Layer 3 outputs and the original input of $P_i$ in the fuzzy rules by multiplying $\overline{w_r}$. The outputs are referred to as consequent parameters.

$$O_r^4 = \overline{w_r} P_{ir} \quad (15)$$

Layer 5: This layer aims to compute the overall output with the sum of all reasoning results from Layer 4.

$$O_r^5 = \sum_r O_r^4 = P_i = \sum_r \overline{w_r} P_{ir} \quad (16)$$

- **Membership Function**

This section describes the triangular membership function utilized in this work; this particular function is depicted in Fig. 4. Each rating level has the corresponding triangular membership function. This membership function is a piecewise-linear function. Throughout the learning process, the membership function is maintained in a fixed state. The following calculation defines the triangular membership function:

$$\mu_{A_{ir}}(x) = \begin{cases} x-(r-1), r-1 \leq x \leq r \\ (r+1)-x, r \leq x \leq r+1 \\ 0, otherwise \end{cases} \text{ for } r = 1, 2, \ldots, 18 \quad (17)$$

where $x = PR_i$ or $x \in [0,19]$

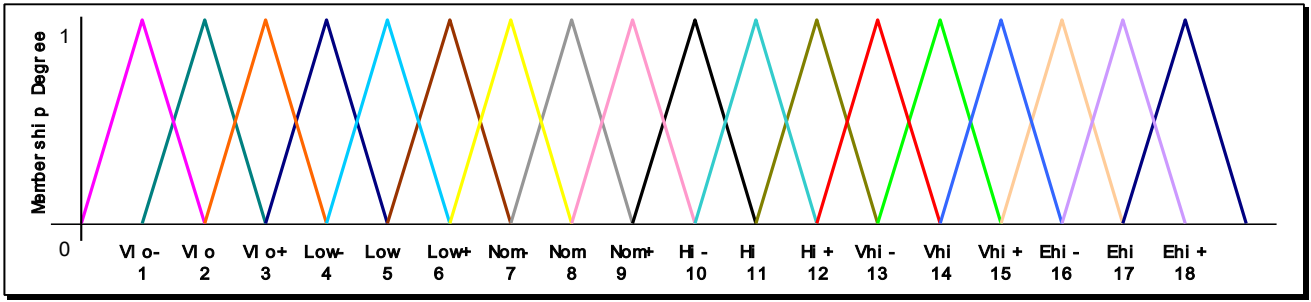

Fig.4. Triangular Membership Function

There are several factors that influenced our selection of the triangular membership function; first, the nature of the NFi outputs was the most crucial reason. Pir is a piecewise-linear interpolation

$$\frac{y-y_0}{y_1-y_0} = \frac{x-x_0}{x_1-x_0}$$

between parameter values ($P_{i1}, \ldots P_{i18}$) of the $i$th parameter, $P_i$. Hence, the selection of the triangular function can be derived from the same results as a linear interpolation. Secondly, one of the purposes of this research is to evaluate the extent to which Huang's proposed soft computing framework can be generalized. Therefore, it was important to use the same membership function as that utilized in Huang's research in order to perform validation with a similar fuzzy logic technique (Huang 2003). Finally, the triangular membership function is easy to calculate.

- **Learning Algorithm**

With ANFIS, there is a two-pass learning cycle: the forward pass and the backward pass. The pass that is selected depends on the trained parameters in ANFIS. In our proposed model, when the error exists between the actual effort and the estimated effort, the outputs are fixed and the inputs are trained. Hence, the backward pass is the type of learning algorithm that this study uses. It is generally a hybrid method of Least Square Estimate (LSE) and Back Propagation, which is calculated using a gradient decent algorithm that minimizes the error. For the learning algorithm, the parameters of the premise part and the consequent part are defined in two sets, as illustrated below:

$$X = \{x_1, x_{2,\ldots}, x_n\}$$
$$= \{PR_1, PR_{2,\ldots}, PR_N, SIBR, Size\} \quad (18)$$
$$P = \{\{P_{11}, P_{21}, \ldots, P_{N1}\}, \{P_{12}, P_{22}, \ldots, P_{N2}\}, \ldots, \{P_{1M}, P_{2M}, \ldots, P_{NM}\}\} \quad (19)$$

where $N = 34$ and $M = 18$; $X$ represents the inputs of the model, which are the rating levels, SIBR and Size; and $P$ is the parameter values of the parameters.

The output of each NF can be defined when substituting (13) and (14) into (16):

$$P_i = f_{NF_i}(P_{i1}, P_{i2}, \ldots, P_{i18}) = \sum_r \overline{w}_r P_{ir} = \sum_{r=1}^{18} \mu_{A_{ir}}(x_i) P_{ir}$$
for $i = 1, 2, \ldots, 34$ \quad (20)

$P_i$ is the weighted sum of inputs $X$ for $PR_i$.

In the section III.B, the equations for the SEER-SEM Effort Estimation are described in detail. The equations (1), (2), (3), (4), and (5) can be re-written as follows with the parameters symbols:

$$Effort = 0.393469 \times P_{34}^{0.4} \times \frac{Size^{1.2}}{2000^{1.2} \times \exp\left(\frac{-3.70945 \times \ln\left(\frac{ctbx}{4.11}\right)}{5 \times P_{10}}\right)^{1.2}} \times ParmAdjustment^{1.2} \quad (21)$$

$ctbx = P_1 \times P_{2-25} \times P_8 \times P_3 \times P_9 \times P_{11}$ \quad (22)

$ParmAdjustment$
$= P_{23-4} \times P_{31-6} \times P_{24-5} \times P_{26-7} \times P_{35-22} \times P_{12} \times \ldots \times P_{21} \times P_{27} \times \ldots \times P_{30} \times P_{33} \times P_{32}$ \quad (23)

Utilizing equations (18) to (21), the proposed neuro-fuzzy model can be written:

$$Effort = f_{NF}(X,P) \tag{24}$$

If there are NN project data points, the inputs and outputs can be presented as $(X_n, E_{acn})$, where n = 1, 2,..., NN, $X_n$ contains 34 parameters as well as SIBR and Size, $E_{aen}$ is the actual effort with $X_n$ inputs for project n. The learning procedure involves adopting the gradient descent method to adjust the parameter values of rating levels that minimizes the error, $E$. According to LSE, the error, $E$, on the output layer is defined as follows:

$$E = \frac{1}{2}\sum_{n=1}^{NN} w_n \left(\frac{E_{en} - E_{acn}}{E_{acn}}\right)^2 \tag{25}$$

where $w_n$ is the weight of project n and $E_{en}$ is the estimation of the output for project n.

$$E_{en} = Effort_n = f_{NF}(X_n, P_n) \tag{26}$$

The following steps are used to perform gradient descent according to the Back Propagation learning algorithm. According to the SEER-SEM effort estimation model presented by equations (21) to (23), the results of the partial derivative of $E_{en}$ with respect to

$P_{ir}$, $\frac{\partial E_{en}}{\partial P_{ir}}$, are different.

$$\frac{\partial E}{\partial P_{ir}} = \sum_{n=1}^{NN} \frac{w_n}{E_{en}^2}(E_{en} - E_{acn})\frac{\partial E_{en}}{\partial P_{ir}} \tag{27}$$

$$\frac{\partial E_{en}}{\partial P_{ir}} = \frac{\partial E_{en}}{\partial P_i}\frac{\partial P_i}{\partial P_{ir}} = \frac{\partial(f_{NF}(X_n,P_n))}{\partial P_i}\frac{\partial P_i}{\partial P_{ir}}$$

for i = 1, 2, ..., 34  (28)

$$\frac{\partial P_i}{\partial P_{ir}} = \frac{\partial(f_{NFi}(P_{ir}))}{\partial P_{ir}} = \frac{\partial(\mu_{A_{ir}}(x_{ir})P_{ir})}{\partial P_{ir}} = \mu_{A_{ir}}(x_i) \tag{29}$$

$\frac{\partial E_{en}}{\partial P_{ir}}$

After    is calculated out, equation (30) is used to calculate the adjusted parameter values.

$$P_{ir}^{l+1} = P_{ir}^{l} - \alpha\frac{\partial E}{\partial P_{ir}} \tag{30}$$

where $\alpha > 0$ is the learning rate and $l$ is the current iteration index.

- **Monotonic Constraints**

A monotonic function is a function that preserves the given order. The parameter values of SEER-SEM are either monotonic increasing or monotonic decreasing. The relationship between the monotonic functions and the rating levels have been accepted by the practitioners as a common sense practice. For instance, the values of ACAP are monotonic decreasing from VLo- to EHi+, which is reasonable because the higher the analysts' capability, the less spent on project efforts. As for TEST, its values are monotonic increasing because the higher test level causes more effort to be spent on projects. After calibrating parameter values by the proposed model, the trained results of these values may contravene the monotonic orders, so that the trained values are changed to a non-monotonic order. For instance, the parameter value of the ACAP rating Hi can be greater than the value of the corresponding rating, EHi. This discrepancy can lead to unreasonable inputs for performing estimation and can impact the overall accuracy. Therefore, monotonic constraints are used by our model in order to maintain consistency with the rating levels.

IV.    EVALUATION

For evaluating the neuro-fuzzy SEER-SEM model, in total, data from 99 studies is collected, including 93 published COCOMO 81 projects and 6 industry studies in the format of COCOMO 87 (Ho 1996; Panlilio-Yap and Ho 2000). An algorithmic estimation model, E = a×Size$^b$ comprises the general form of COCOMO and SEER-SEM (Fischman, McRitchie, and Galorath 2005; Jensen, Putnam, and Roetzheim 2006). Specifically, this model enables us to use the COCOMO database for evaluating the proposed SEER-SEM model in spite of the difference between COCOMO and SEER-SEM. In fact, various studies have revealed the similar estimation performances of COCOMO and SEER-SEM (Madachy, Boehm, and Wu 2006; USC Center for Software Engineering 2006).

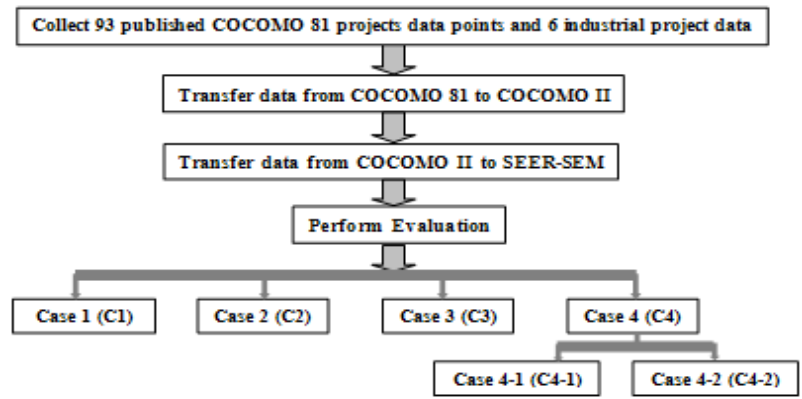

Fig.5. Main Evaluation Steps.

Fig. 5 shows the main steps of our evaluation. First, in order to use both published COCOMO 81 and industrial project data in the evaluation, the information was translated into the corresponding format of SEER-SEM data. Second, there are four cases for evaluating the prediction performance of our neuro-fuzzy model.

1) *Performance Evaluation Metrics*

The following evaluation metrics are adapted to assess and evaluate the performance of the effort estimation models.

- **Relative Error (RE)**

$$RE = \frac{(EstimationEffort - ActualEffort)}{ActualEffort}$$

The RE is used to calculate the estimation accuracy.

- **Magnitude of Relative Error (MRE)**

$$MRE = \frac{|EstimationEffort - ActualEffort|}{ActualEffort}$$

- **Mean Magnitude of Relative Error (MMRE)**

$$MMRE = \frac{\left(\sum_{i=1}^{n} MRE_i\right)}{n}$$

The MMRE calculates the mean for the sum of the MRE of n projects. Specifically, it is used to evaluate the prediction performance of an estimation model.

- **Prediction Level (PRED)**

$$PRED(L) = \frac{k}{n}$$

where L is the maximum MRE of a selected range, n is the total number of projects, and k is **number of projects** in a set of n projects whose MRE <= L. PRED calculates the ratio of projects' MREs that falls into the selected range (L) out of the total projects.
(e.g. n = 100, k =80, where L= MRE <= 30%: PRED(30%) = 80/100 = 80%)

2) *Dataset*

There are two major steps in transferring data from COCOMO 81 to SEER-SEM: first, information is converted from COCOMO 81 to COCOMO II and then from COCOMO II to SEER-SEM. The main guidelines are referred to (Madachy, Boehm, and Wu 2006; Reifer, Boehm, and Chulani 1999). In the method of the second step, 20 of the 34 SEER-SEM technical parameters can be directly mapped to 14 COCOMO II cost drivers and 1 scale factors, 1 COCOMO 81 cost driver, and 2 COCOMO 87 cost drivers. The remainder of the SEER-SEM parameters cannot be transferred to the COCOMO model, and as a result, they are set up as nominal in SEER-SEM. After transferring 93 COCOMO 81 project data points, the estimation performance with transferred data are evaluated with the estimation performance metrics. Table 1 presents the details of the prediction performance of COCOMO 81, COCOMO II, and SEER-SEM.

Table 1. Estimation Performance with Transferred Data.

|  | Cocomo 81 | Cocomo II | Seer-sem |
|---|---|---|---|
| Mmre (%) | 56.46 | 48.63 | 84.39 |
| Pred(20%) | 36.56 | 37.63 | 36.56 |
| Pred(30%) | 51.61 | 54.84 | 45.16 |
| Pred(50%) | 76.34 | 78.49 | 56.99 |
| Pred(100%) | 92.47 | 94.62 | 81.72 |
| # of Outliers | 22 | 20 | 39 |

The data transferring from COCOMO 81 to COCOMO II keeps the very close performance with little improvement when doing COCOMO II estimation with the transferred data. The transferring from COCOMO II to SEER-SEM causes the MMRE decreasing and the outliers increasing. Most of the new outliers come from the embedded projects whose MREs are lower than 50% before being transferred to SEER-SEM. The PRED is still stable and there is not a huge change. Overall, transferring from COCOMO 81 to SEER-SEM is feasible for our evaluation, especially when the actual project data in the format of SEER-SEM are difficult to obtain. We use the online calculator of the USC Center for Software Engineering to perform COCOMO 81 and COCOMO II estimation. We do SEER-SEM effort estimation by two methods. One is performed by the SEEM-SEM tool (SEER-SEM for Software 7.3) which is offered by GAI, and the other is done manually by Microsoft Excel with the equations of SEER-SEM effort estimation model as presented in the section III.B. The SEER-SEM effort estimation model is also implemented as part of our research because it is part of our proposed model. The estimation performance by the SEER-SEM tool and Excel are very close. This is a way to make sure the algorithm of SEER-SEM effort estimation presented in this paper to be correct. We select the results done manually to avoid the impact from other parameters settings in the SEER-SEM tool.

The dataset of 6 industrial project data points is from the COCOMO 87 model, which is slightly different than COCOMO 81, as the effort multipliers RUSE, VMVH (Host Volatility), and VMVT (Target Volatility) are not used in COCOMO 81. However, RUSE can be transferred to COCOMO II directly because it is one of the COCOMO II cost drivers, and VMVH and VMVT can be transferred to the SEER-SEM parameters DSVL and TSVL. The rest of COCOMO 87 cost drivers are matched to the corresponding cost drivers of COCOMO 81. Then, they are transferred to COCOMO II and SEER-SEM.

3) *Evaluation Cases*

After transferring the data, we conducted four main case studies to evaluate our model. These cases, which used different datasets from 93 projects, were utilized to perform training on the parameter values. The 93 project data points and the 6 industrial project data points were adopted for testing purposes. The original SEER-SEM parameter values are trained in each case. The learned parameter values of the four cases are different. This reason causes the prediction performance difference amongst the cases and the SEER-SEM. In order to assess the prediction performance of the neuro-fuzzy model, we compared SEER-SEM effort estimation model with our framework. Several performance metrics were used for the analysis of each case, including MRE, MMRE, and PRED. Accordingly, Table 2 presents the MMRE results from Cases 1 to 4, and Table 3 shows the MMRE results of the industrial project data points. Table 4 shows the PRED results of Cases 1, 2, and 3. The PRED results of Case 4 are presented in Table 5. In the tables presenting the analysis results, we have included a column named "Change", which is used to indicate the performance difference between SEER-SEM effort estimation model and our neuro-fuzzy model. For the MMRE, the prediction performance improves as the value becomes closer to zero; therefore, if the change for these performance metrics is a negative value, the MMRE for the neuro-fuzzy model is improved in comparison with SEER-SEM. Additionally, the "PRED(L)" in Table 4 represent the prediction level of the selected range, referring to the definition presented in the section IV.A; a higher prediction level indicates a greater level of performance for PRED. For PRED, a negative value for the "Change" indicates that our model shows a decreased level of performance as compared to SEER-SEM. Finally, the results for both MMRE and PRED are shown in a percentage format.

Table 2. MMRE of 93 Published Data Points.

| Case ID | SEER-SEM | Validation | Change |
|---|---|---|---|
| C1 | 84.39 | 61.05 | -23.35 |
| C2 | 84.39 | 59.11 | -25.28 |
| C3 | 84.39 | 59.07 | -25.32 |
| C4-1 | 50.49 | 39.51 | -10.98 |
| C4-2 | 42.05 | 29.01 | -13.04 |

Table 3. MMRE of Industrial Project Data Points.

| Case ID | MMRE (%) | | |
|---|---|---|---|
| | SEER-SEM | Industrial Average | Change |
| C1 | 37.54 | 35.54 | -2 |
| C2 | 37.54 | 47.57 | 10.03 |
| C3 | 37.54 | 47.16 | 9.62 |
| C4-1 | 37.54 | 33.20 | -4.34 |
| C4-2 | 37.54 | 30.39 | -7.15 |

Table 4. PRED of Cases 1, 2 and 3.

| | SEER-SEM | Neuro-Fuzzy Model | | | | | |
|---|---|---|---|---|---|---|---|
| PRED(L) | PRED (%) | C1 | | C2 | | C3 | |
| | | PRED (%) | Change | PRED (%) | Change | PRED (%) | Change |
| PRED(20%) | 36.65 | 29.03 | -7.62 | 15.05 | -21.6 | 15.05 | -21.6 |
| PRED(30%) | 45.16 | 37.63 | -7.53 | 18.28 | -26.88 | 18.28 | -26.88 |
| PRED(50%) | 56.99 | 64.52 | 7.53 | 36.56 | -20.43 | 38.71 | -18.28 |
| PRED(100%) | 81.72 | 92.47 | 10.75 | 97.85 | 16.13 | 97.85 | 16.13 |

*Case 1 (C1): Learning with project data points excluding all outliers*

This case involved training the parameters of projects where the MREs are lower than or equal to 50%. There are 54 projects that meet this requirement. Since we wanted to perform learning without any impact from the outliers, the learning was done with 54 project data points, while 93 pieces of project data and the 6 industrial project data points were used for testing. When using the neuro-fuzzy model, the MMRE decreased from 84.39% to 61.05%, with an overall improvement of 23.35%. After testing data from the 93 projects, we used the 6 industrial project data points to perform testing. The results of this evaluation present the same tendency as the testing results with the 93 project data points: the MMRE of the neuro-fuzzy model is lower than the MMRE of SEER-SEM by 2%. With the neuro-fuzzy model, PRED(20%) and PRED(30%) decreased by 7.62% and 7.53% in comparison to the same values using SEER-SEM; however, PRED(50%) and PRED(100%) improved with the neuro-fuzzy model by a factor of 7.53% and 10.75% respectively, which indicates that the MRE of the neuro-fuzzy model, in comparison with that of SEER-SEM, contained more outliers that were less than 100% or 50%. Furthermore, the MMRE was significantly improved with the neuro-fuzzy model due to the increase of outliers that were less than 100%. By integrating the results from the MMRE, PRED, and the industrial project data points, this calibration demonstrates that the neuro-fuzzy model has the ability to reduce large MREs.

*Case 2 (C2): Learning with all project data including all outliers*

In Case 2, we used the data points from all 93 projects to calibrate the neuro-fuzzy model without removing the 39 outliers. The testing was performed with the same project dataset used in the training and with the 6 industrial project data points. In comparison to Case 1, this test attempted to ascertain the prediction performance when the learning involved the outliers as well as the effects of the outliers on the calibration. the MMRE using SEER-SEM comparison to the MMRE using SEER-SEM. Nevertheless, the industrial project data points caused the MMRE to worsen with the neuro-fuzzy model by 10.03%. The results of PRED demonstrate that PRED(20%), PRED(30%), and PRED(50%) decreased by more than 20%, while PRED(100%) increased by 16.13% with the neuro-fuzzy model. Moreover, these results also indicate that the neuro-fuzzy model is effective for improving the MREs that are greater than 100%. As a result, the MMRE in all of the datasets are improved when the neuro-fuzzy model is utilized. In Cases 1 and 2, the results of PRED and the 6 industrial project data points show that the neuro-fuzzy model causes large increases in small MREs while reducing large MREs. Hence, the decrease of large MREs leads to the overall improvement of the MMRE, thus showing the effectiveness of the neuro-fuzzy model.

*Case 3 (C3): Learning with project data excluding part of outliers*

After training, which included and then excluded all of the outliers, Case 3 calibrated the neuro-fuzzy model by removing the top 12 of 39 outliers where the MRE is more than 150%. In this case, 87 project data points are used to perform training, and the 93 project data points and the 6 industrial project data points are used for testing. The results of Case 3 are almost identical to the results of MMRE and PRED as demonstrated in Case 2. Specifically, for the neuro-fuzzy model, the MMRE of industrial project data points is worsened by 9.62%. Overall, as compared to Case 2, calibration excluding the top 12 outliers does not make a significant difference in the performance of the model.

*Case 4 (C4): Learning with part of project data points*

In the previous three cases, all data points from the 93 projects were used for testing. However, in Case 4, we used part of this dataset to calibrate the neuro-fuzzy model, and the rest of the data points, along with the 6 industrial project data points, were used for testing. The objective of this case was to determine the impact of the training dataset size on the calibration results. Table 2, Table 3, and Table 5 present the results.

*Case 4 -1 (C4-1):*

*Learning with 75% of project data points and testing with 25% of project data points*

This sub-case performed training with 75% of the 93 project data points and testing with the remaining 25% of these points. The project numbers for the training data points ranged from 24 to 93, while those for the testing points ranged from 1 to 23 and also included the 6 industrial project data points. To analyze the results, we compared the performance of SEER-SEM to that of the neuro-fuzzy model for Projects 1 to 23. In this case, the neuro-fuzzy model improved the MMRE by 10.98%. Furthermore, PRED(30%) and PRED(100%) with our model improved by 4.35% and 8.70% respectively. Finally, with the neuro-fuzzy model, the MREs of all 23 project data points were within 100%. In this case, the testing results of the industrial project data points are improved from the previous tests by 4.34%. These results demonstrate the effective performance of the neuro-fuzzy model in reducing large MREs.

- *Case 4 -2 (C4-2):*
  *Learning with 50% of project data points and testing with 50% of project data points*

Case 4-2 divided the 93 project data points into two subsets. The first subset included 46 project data points that are numbered from 1 to 46 and were used to perform testing. On the other hand, the second subset contained 47 project data points, numbered from 47 to 93, which were used to train the neuro-fuzzy model. In comparison to Case 4-1, this test contains fewer training data points and more testing data points. Accordingly, we analyzed the performance results of the 46 project data points as estimated by both SEER-SEM and the neuro-fuzzy model. In this case, the MMRE improved by 13.04% when using the neuro-fuzzy model. Specifically, the results of PRED showed improvement from those in Case 4-1; not only were the MREs of all 46 project data points within 100%, but the MREs of most project data points were also less than 50%. Furthermore, in the testing that involved the 6 industrial project data points, the results were better than those in Case 4-1. Using the neuro-fuzzy approach, the MMRE of the 6 industrial project data points improved by 7.15%, which was the greatest improvement among all of the cases in this study.

Table 5. Case 4 PRED Results.

| PRED (L) | C4-1 PRED (%) | | | C4-2 PRED (%) | | |
|---|---|---|---|---|---|---|
| | SEER-SEM | Neuro-Fuzzy Model | Change | SEER-SEM | Neuro-Fuzzy Model | Change |
| PRED(20%) | 39.13 | 34.78 | -4.35 | 50.00 | 43.48 | -6.52 |
| PRED(30%) | 47.83 | 52.17 | 4.35 | 63.04 | 56.52 | -6.52 |
| PRED(50%) | 65.22 | 60.87 | -4.35 | 73.91 | 76.09 | 2.17 |
| PRED(100%) | 91.30 | 100 | 8.70 | 91.30 | 100 | 8.7 |

4) EVALUATION SUMMARY

In this section, we summarize the evaluation results by comparing the analysis of all of the cases as presented in the previous sections. Fig. 6 shows the validation summary for the mmre across all of the cases. Specifically, the mmre improves in all of the cases, with the greatest improvement being over 25%.

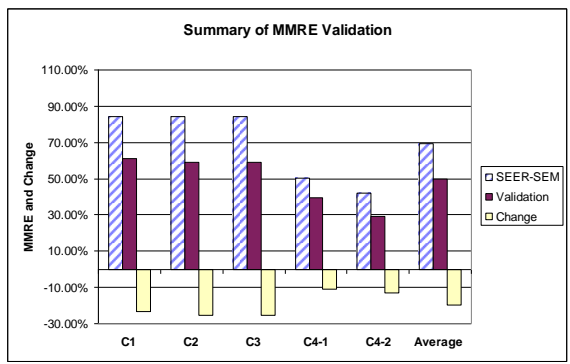

Fig.6. Summary of MMRE Validation.

Table 6 illustrates the PRED averages for SEER-SEM in all of the cases, and Fig. 7 shows the PRED averages for all of the cases using the neuro-fuzzy model. Compared to the PREDs from SEER-SEM, the averages of PRED(20%), PRED(30%), and PRED(50%) with the neuro-fuzzy model do not show improvement. However, the average of PRED(100%) is increased by 12.14%, which indicates that the neuro-fuzzy model improves the performance of the MMRE by reducing the large MREs.

Table 5. Summary of PRED Average.

| | SEER-SEM | Average of Validation | Change |
|---|---|---|---|
| PRED(20%) | 39.76% | 27.48% | -12.28% |
| PRED(30%) | 49.27% | 36.46% | -12.81% |
| PRED(50%) | 62.02% | 55.35% | -6.67% |
| PRED(100%) | 85.55% | 97.69% | 12.14% |

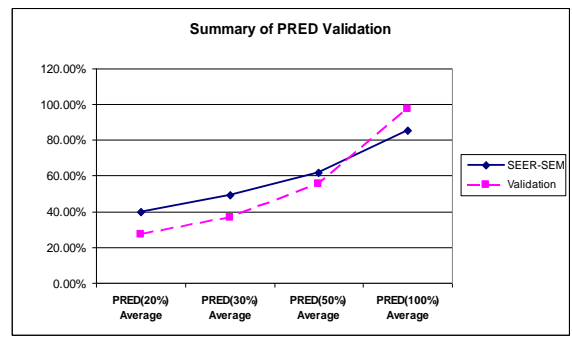

Fig.7. Summary of PRED Validation

Fig. 8 presents the MMREs of industrial project data points from all of the cases. The MMRE from Cases 1 and 4 demonstrate an improvement of no more than 7.15%. The calibrations with the outliers in Cases 2 and 3 lower the prediction performance of these two cases. Thus, for the neuro-fuzzy model, the improvement of the MMRE of industrial projects is minimal.

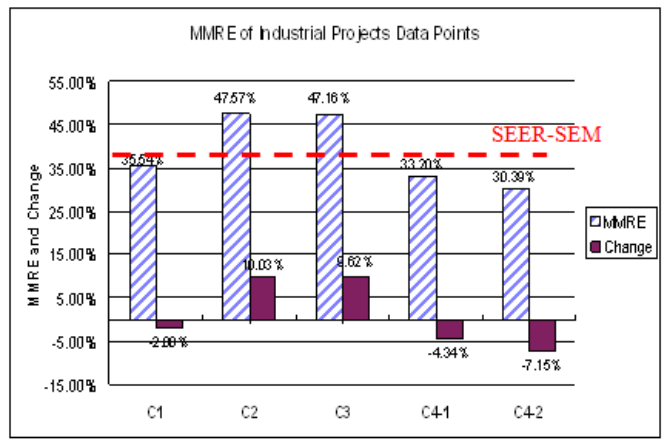

Fig.6. MMRE of Industrial Project Data Points.

## V. Conclusion

Overall, our research demonstrates that combining the neuro-fuzzy model with the SEER-SEM effort estimation model produces unique characteristics and performance improvements. Effort estimation using this framework is a good reference for the other popular estimation algorithmic models. The neuro-fuzzy features of the model provide our neuro-fuzzy SEER-SEM model with the advantages of strong adaptability with the capability of learning, less sensitivity for imprecise and uncertain inputs, easy to be understood and implemented, strong knowledge integration, and high transparency.

Four main contributions are provided by this study:

a) ANFIS is a popular neuro-fuzzy system with the advantages of neural network and fuzzy logic techniques, especially the ability of learning. The proposed neuro-fuzzy model can successfully manage the nonlinear and complex relationship between the inputs and outputs and it is able to handle input uncertainty from the data.

b) The involvement of fuzzy logic techniques improves the knowledge integration of our proposed model. Fuzzy logic has the ability to map linguistic terms to variables. Accordingly, the inputs of our model are not limited to linguistic terms and can also work with numerical values. The defined fuzzy rules are an effective method for obtaining the experts' understanding and experience to produce more reasonable inputs.

c) There are two techniques introduced in this research: the triangular membership function and the monotonic constraint. Triangular Membership Functions are utilized to translate parameter values to membership values. Furthermore, monotonic constraints are used in order to preserve the given order and maintain consistency for the rating values of the SEER-SEM parameters. These techniques provide a good generalization for the proposed estimation model.

d) This research proves that the proposed neuro-fuzzy structure can be used with other algorithmic models besides the COCOMO model and presents further evidence that the general soft computing framework can work effective with various algorithmic models. The evaluation results indicate that estimation with our proposed neuro-fuzzy model containing SEER-SEM is more efficient than the estimation results that only use SEER-SEM effort estimation model. Specifically, in all four cases, the MMREs of our proposed model are improved over the ones where only SEER-SEM effort estimation model is used, and there is more than a 20% decrease as compared to SEER-SEM. According to these results, it is apparent that the neuro-fuzzy technology improves the prediction accuracy when it is combined with the SEER-SEM effort estimation model, especially when reducing the outliers of MRE >100%.

Although several studies have already attempted to improve the general soft computing framework, there is still room for future work. First, the algorithm of the SEER-SEM effort estimation model is more complex than that of the COCOMO model. Prior research that combines neuro-fuzzy techniques with the COCOMO model demonstrates greater improvements in the prediction performance. Hence, the proposed general soft computing framework should be evaluated with other complex algorithms. Secondly, the datasets in our research are not from the original projects whose estimations are performed by SEER-SEM. When the SEER-SEM estimation datasets are available, more cases can be completed effectively for evaluating the performance of the neuro-fuzzy model.